\documentstyle[pra,aps,twocolumn]{revtex}
\input psfig.tex

\begin{document}
\author{T. Pellizzari}
\title{Quantum Networking with Optical Fibres}
\address{Clarendon Laboratory, Departement of Physics, University of Oxford, \\Parks
Road, OX2 6RX Oxford, England.}
\maketitle

\begin{abstract}
I propose a scheme which allows for reliable transfer of quantum 
information 
between two atoms
via an optical
fibre in the presence of decoherence. The scheme is based on performing an
adiabatic passage through two cavities which remain in their respective
vacuum states during the whole operation. 
The scheme may be
useful for networking several ion--trap quantum computers, thereby
increasing the number of quantum bits involved in a computation.
\end{abstract}


\narrowtext

\vspace{0.3cm}
The possibility of reliable spacial transport of quantum states is
of crucial importance to quantum communication. During the past 
few years rapid progress has been made in 
using optical fibres for quantum communication on the single
photon level \cite{crypto1}. 
This is needed for quantum 
cryptography \cite{crypto1,crypto2} and quantum teleportation 
\cite{qtele}. 
However, the 
spatial exchange of quantum information between quantum registers that 
have undergone local quantum processing has not yet been demonstrated
experimentally. Quantum communication
between locally distinct nodes of a quantum network will be essential to
overcome small--scale quantum computing \cite{qcomp}. 

A promising model for storage and local processing 
of quantum information is the ion--trap quantum computer, in which the 
quantum bits are stored in stable ground states or in
long--lived metastable states \cite{ciraczoller}. First experimental results have
already been reported \cite{nistboulder}. 
However,  technology imposes an upper limit on 
the number of ions, and thus quantum bits, which can be used in a single
ion--trap quantum computer. To overcome this limit,
a network of several ion trap quantum computers could be set up. 
But while ions are superb for storing quantum information, 
it seems more feasible to mediate quantum states by photons carried by optical
fibres.
\begin{figure}
\center{\psfig{file=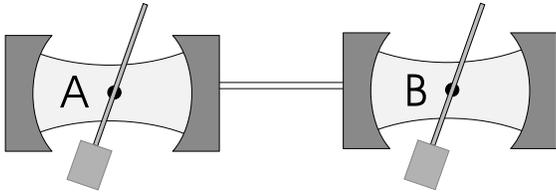,width=2.9in} }
\caption{Representation of the scheme. The atoms (quantum bits)
are coupled to cavities and are manipulated by lasers. The cavities 
are connected via an optical fibre.}
\end{figure}

This problem has been recently addressed (to my knowledge, for the first 
time)
by Cirac {\it et al.} \cite{symmetricwavepackets}. 
They propose a scheme to transmit quantum bits by 
tailoring {\it time--symmetric photon wave--packets}. In the 
present letter, I will pursue a different approach which 
is based on an {\it adiabatic passage via photonic dark states}.
A representation of the scheme is depicted in 
Fig.~1. 

The quantum bit to be transferred is initially stored in
atom A, and atom B is prepared in a predefined
state. Atoms A and B may be part of ion--trap quantum 
computers. Initially both cavities and the fibre are in the
vacuum state. By appropriate design of 
laser pulses as described below we may 
swap the states of atom A and B.
Below I will demonstrate that (ideally)
the two cavities 
will never have a photon number different from zero.

The scheme has two distinctive features: Firstly, it is 
insensitive to losses from the cavities into other than the
fibre modes. This is due to the fact that the cavities are
never populated. As a result, the coupling between the cavities
and the fibre mode does not need to be perfect; Secondly,
it does not require a precise control
of the pulse shape, duration and intensity of the manipulating
laser pulses is not required as long as some ``global'' 
(namely adiabaticity--) conditions are met. This implies that 
the Rabi frequencies need not be known in order to perform 
the transfer successfully.  

The individual atom--cavity systems are described as follows.
The atoms are modelled by three--level $\Lambda$--systems
with two ground states $|a_0\rangle$, $|a_1\rangle$ and one
excited state $|b\rangle$ as depicted in Fig.~2a. 
\begin{figure}
\center{\psfig{file=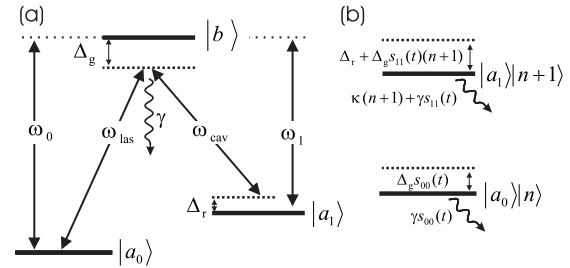,width=2.9in} }
\caption{(a) Atomic level scheme. See text for definition of the symbols. 
(b) Level scheme of the combined
atom--cavity system after adiabatic 
elimination. The first parameter in the ket 
represents an atomic ground state whereas the second parameter denotes 
a cavity mode Fock state.}
\end{figure}
The frequency of the transition $|a_i\rangle-|b\rangle$
is denoted by $\omega_i$, where $i=0,1$. The excited state
$|b\rangle$ spontaneously decays with a rate $\gamma$.
The cavity
is modelled by a single, quantized mode with frequency 
$\omega_{\rm cav}$ coupled to the transition $|a_1\rangle-|b\rangle$ 
with coupling strength $g$. 
The cavity is coupled to an optical fibre as described below.
In addition, we include a loss rate $\kappa$ of the cavity.
This decay rate includes any undesired loss mechanisms 
such as absorption of cavity photons in the 
mirrors and coupling to other than the fibre modes.
The transition $|a_0\rangle-|b\rangle$
is coupled to a laser described by a c--number coherent field with 
frequency $\omega_{\rm las}$. The corresponding 
time--dependent Rabi frequency is denoted by $\Omega(t)$.
According to continuous measurement theory the evolution of
a system in the absence of quantum jumps is determined by
an effective, non--Hermitian Hamiltonian \cite{stochschroedinger}. 
In the present context
we find:
\begin{eqnarray} \label{Heff}
\nonumber
H_{\rm eff} 	&=& 	H_0 +  H_{\rm int}\\
\nonumber
H_0 		&=&	(-\Delta_{\rm g} - i\gamma) |b\rangle \langle b| 
			+ \Delta_{\rm r} |a_1 \rangle \langle a_1|
			+ \kappa a^{\dagger} a \\
H_{\rm int}	&=&	\Omega(t) |b\rangle \langle a_0| +
			g a |b\rangle \langle a_1| + H.c.
\end{eqnarray}
Eq.~(\ref{Heff}) is written in a rotating frame. The symbols $a$ and 
$a^{\dagger}$ denote the annihilation and creation operators of the
cavity mode. Here I have introduced two detunings,
namely the ``global'' detuning $\Delta_{\rm g} = \omega_{\rm las}
-\omega_0$ and the ``Raman'' detuning $\Delta_{\rm r} = 
(\omega_{\rm cav}-\omega_{\rm las}) - (\omega_0 - \omega_1)$.

We are interested in the low saturation regime where we can 
adiabatically eliminate the excited atomic state. Below
the conditions that are required to make this approximation
meaningful are given. 
By applying standard quantum optical techniques \cite{quantumnoise} 
the adiabatically eliminated effective Hamiltonian reads:
\begin{eqnarray} \label{adelHeff}
\nonumber
\tilde{H}_{\rm eff} 	&=& 	\tilde{H_0} +  \tilde{H_1} + 
				\tilde{H}_{\rm int}
\\ \nonumber
\tilde{H}_0		&=&	+ \left( \Delta_{\rm g} s_{00}(t)
				-i\gamma s_{00}(t) \right)
				|a_0\rangle \langle a_0|
\\ \nonumber
 \tilde{H_1}		&=&	\left( \Delta_{\rm r} + 
				\Delta_g s_{11}(t) a^{\dagger} a 
				-i \kappa  a^{\dagger} a -i\gamma
				s_{11}(t)\right) |a_1\rangle \langle a_1|
\\ \nonumber
\tilde{H}_{\rm int} 	&=&	\left( \Delta_g - i \gamma \right) s_{01}(t) a
				|a_0\rangle \langle a_1| +
\\ 
			&&	\left( \Delta_g - i \gamma \right) s_{10}(t) 
				a^{\dagger}
                                |a_1\rangle \langle a_0|.
\end{eqnarray}
Here I have introduced the saturation parameters 
\begin{eqnarray}
\nonumber
s_{00}(t) &=& \frac{|\Omega(t)|^2}{\Delta_{\rm g}+\gamma^2}  
\quad\quad
s_{11}(t) = \frac{|g|^2}{\Delta_{\rm g}^2+\gamma^2} \\
s_{01}(t) &=& s_{10}^*(t) = \frac{\Omega(t)g^*}{\Delta_{\rm g}^2+\gamma^2}.
\end{eqnarray}
Adiabatic elimination is applicable if the following 
inequalities hold:
\[
	s_{ii}(t) \ll 1, \quad i=1,2.
\]
The detunings, light--shifts, and decay rates of the two ground states 
are shown in Fig.~2b.
In Eq.~(\ref{adelHeff}), $\tilde{H_0}$ and $\tilde{H}_1$ contain 
all the terms which
are diagonal in the atomic basis. Both ground states
undergo light shifts and damping. Later we will
find that the light shift of state $|a_0\rangle$ gives
rise to undesired effects which will make 
compensate necessary. $\tilde{H}_{\rm int}$ describes Rabi 
oscillations between the ground states and contains
dissipative terms as well. The decay terms are of course
unwanted. However, it will be shown in the following how 
they can be avoided.

The next step is to consider two atom--cavity systems as described 
above connected by an optical fibre. 
The free Hamiltonian of the fibre and the interaction Hamiltonian 
with the two atom--cavity systems is assumed to take the following form:
\begin{equation} \label{fiber}
H_{\rm fib} = \sum_k \Delta_k f_k^\dagger  f_k +
\left\{
\nu \sum_k
\left(
a_A + (-1)^k a_B
\right) f_k + H.c.
\right\}
\end{equation}
Eq.~(\ref{fiber}) is written in a frame rotating at the 
cavity frequency. By $f_k(f_k{\dagger})$ I denote the annihilation
(creation) operator of the $k$th fibre mode. $\Delta_k$ is the frequency
difference between the $k$th fibre mode and the cavity mode, and
$\nu$ is the coupling strength between the fibre modes and the cavity.
For simplicity it is assumed that in the frequency range where 
the coupling is significant the coupling strength is constant. 
Here, and in the remainder of this letter, the subscripts (or superscripts)
$A$ and $B$ distinguish the two atom--cavity sub--systems.
Note that the fibre is assumed to be lossless which is of course 
unrealistic for long transmission distances. However, the present 
scheme is designed for short distances for which 
this assumption seems appropriate.

The goal is to achieve quantum state swapping. 
In this model energy eigenstates of the system and
logical values are identified as follows:
\[
|i\rangle|j\rangle \equiv |a_i,0\rangle_A |a_j,0\rangle_B |{\rm vac}\rangle 
\quad\quad i=0,1
\]
Here the first and the second ket on the right hand side refer to 
the atom--cavity subsystem $A$ and $B$. The first parameter
in each ket denotes the atomic ground state, while the second 
represents a Fock state of the respective cavity modes. The 
third ket denotes the subsystem of the fibre--modes all being in the
vacuum state. Below I will demonstrate how the following 
operation can be performed:
\begin{equation} \label{transformation}
\left( \alpha |0\rangle + \beta  |1\rangle \right)|1\rangle 
\longrightarrow
|1\rangle \left( \alpha |0\rangle + \beta  |1\rangle \right).
\end{equation}
Heree $\alpha$ and $\beta$ are arbitrary (in general unknown) 
complex amplitudes.
The result of the process by starting with the second quantum bit 
in state $|0\rangle$ will be undefined.

The primary objective is to perform the desired quantum state 
transfer with as little disturbing influence from dissipative
processes as possible. In the present context these are spontaneous
emissions from the atomic excited states (via optical pumping)
and the decay of the cavity
mode into other than the fibre--modes. These two decay mechanisms 
are dealt with as follows: 
(i) 
The optical pumping rate is the 
product of the saturation parameter $s_{ii}(t)$ and the spontaneous
decay rate $\gamma$, whereas the effective Rabi frequency 
is the product of $s_{01}(t)$ and the detuning $\Delta_g$.
Therefore, by increasing the detuning and simultaneously increasing 
the Rabi frequency the optical pumping rate can be made arbitrarily
small while maintaining the effective Rabi frequency constant.
(ii) Undesired loss of cavity photons is avoided by performing
the process as an adiabatic passage through a dark state
of both cavities. In other words, the photon number of two cavity modes 
will not differ significantly from zero throughout the whole
process. In the following paragraphs I will elaborate 
on the details of this scheme.

Adiabatic passage \cite{Ur-AP} has a variety of applications in the 
context of atomic physics. For example, it can be
used for transferring population between atomic levels unaffected
by spontaneous emission \cite{BergmannAP}. 
A beam splitter for atoms based on adiabatic passage
has been proposed
by Marte {\it et al.}  
and realized experimentally by several groups \cite{MarteAP}.
Moreover, in cavity QED quantum state sythesis \cite{ScottAP} and 
quantum computing \cite{TomAP} based on adiabatic passage has been proposed.

Since spontaneous emission can be dealt with in the aforementioned
way we will set $\gamma=0$ in the following. The present scheme 
is based on the fact that for the Hamiltonian $H_{\rm part}=\tilde{H}_1+\tilde{H}_{\rm int}+
\tilde{H}_{\rm fib}$ given in Eqs.~(\ref{adelHeff}) and (\ref{fiber})
dark states with respect to the two cavity modes exists.
This requires that a fibre mode exists that is resonant with the cavity mode,
i.e. that a detuning $\Delta_k=0$ for a certain $k$ exists.
Note that this is not the full Hamiltonian of the system since 
$\tilde{H}_0$ is missing. It turns out that $\tilde{H}_0$
deteriorates the dark state and thus the performance of the scheme.
Below it will be shown how this effect can be avoided. 
The two relevant dark states of the partial 
Hamiltonian $H_{\rm part}$ read thus:
\begin{eqnarray} \label{darkstate} 
\nonumber
|\Psi^D_0\rangle &\propto&
	\nu \Delta s_{10}^B(t) 
	|a_0,0\rangle_A |a_1,0\rangle_B |\rm vac\rangle  
	\\ \nonumber &&
	- \Delta^2 s_{10}^A(t)s_{10}^B(t) 
	|a_1,0\rangle_A |a_1,0\rangle_B
	|1_{k_0}\rangle	
	\\ \nonumber &&
	+\nu \Delta s_{10}^A(t) 
	|a_1,0\rangle_A |a_0,0\rangle_B |\rm vac\rangle  
\\
|\Psi^D_1\rangle &=& |a_1,0\rangle_A |a_1,0\rangle_B |\rm vac\rangle.
\end{eqnarray}
By $|1_{k_0}\rangle$ I denote the fibre state corresponding to 
fibre mode $k_0$ being in a one photon Fock state and all the
others in the vacuum state. The index $k_0$ corresponds to the fibre mode
for which $\Delta_{k_0}=0$.
These states are eigenstates of $H_{\rm part}$ and do not contain
excited states of the cavity mode. In the first dark state 
$|\Psi^D_0\rangle$
the cavity modes are not populated due to 
destructive quantum interference, whereas
the second dark state $|\Psi^D_1\rangle$ is decoupled from the laser 
interaction in a trivial way.

The central idea is to use this dark state for adiabatic passage. 
If the system is prepared in the above dark state (\ref{darkstate})
and the laser intensities are changed slowly (i.e. adiabatically) 
no other eigenstates of $H_{\rm part}$ will be populated. Therefore, throughout 
the whole process the two cavity modes will not be populated.
In our particular case we initially have an unknown 
quantum superposition prepared in subsystem $A$ as given in 
Eq.~(\ref{transformation}). This initial state can in fact 
be written as a superposition of the two dark states
 $|\Psi^D_{0,1}\rangle$ provided that $s_{10}^B(t)\gg s_{10}^A(t)$:
\[
\left( \alpha |0\rangle + \beta |1\rangle \right)|1\rangle
=  \alpha |\Psi^D_0\rangle + \beta|\Psi^D_1\rangle
\]
In practice this means that at the beginning of the transfer the laser
which acts on the atom in subsystem $B$ is switched on first.
During the transfer the laser intensities are changed 
such that at the end the inequality $s_{10}^B(t)\ll s_{10}^A(t)$
holds. If the change is carried out slowly enough 
the system at the end is still the above superposition 
of $|\Psi^D_0\rangle$ and $|\Psi^D_1\rangle$. However, this quantum state 
now corresponds to the desired final state of the transfer, i.e.
\[
\alpha |\Psi^D_0\rangle + \beta|\Psi^D_1\rangle
=|1\rangle\left( \alpha |0\rangle + \beta |0\rangle \right).
\]
An important feature of the scheme is that the 
details of the laser pulses are not important as long
as the process is carried out adiabatically.

As mentioned above, the Hamiltonian ${H}_{\rm part}$
is not the total Hamiltonian of the system because 
it does not contain $\tilde{H}_0$ specified in Eq.~(\ref{adelHeff}).
This Hamiltonian contains light shift of the state 
$|a_0\rangle$ which will destroy the dark state
Eq.~(\ref{darkstate}). However, this light shift can be 
compensated for quite straightforwardly by using a second
laser which couples the atomic level $|a_0\rangle$ 
non--resonantly with an additional level further 
up in the atomic level scheme. The intensity of this
laser is chosen such that the total light shift 
of $|a_0\rangle$ adds up to zero.

In the remainder of this letter I will present numerical 
results in order to evaluate the performance of the 
scheme. 
Note that the coupling strength 
$\nu$ will decrease when the length of the fibre is increased.
Therefore, it is advantageous to introduce a coupling strength 
per unit length defined as:
\[
\alpha=\nu\sqrt{L},
\]
where $L$ is the length of the fibre. We shall express all the 
other parameters in units of $\alpha$, and a 
unit length $L_0$. 
The order of magnitude of $\alpha$ in terms of known quantities
can be estimated as follows.
Suppose the fibre is infinitely long and the decay rate 
of the single cavity mode into the continuum of fibre modes
is $\kappa_f$. As a result of the quantum fluctuation--dissipation
theorem the coupling strength between the cavity mode and the 
output--field of the fibre modes is $g_f=\sqrt{2\kappa_f}$ \cite{quantumnoise}.
Now we estimate the coupling of the cavity mode to the individual 
fibre modes provided that $L$ is finite. Let us first estimate the 
number of fibre modes which are coupled to the cavity 
mode. For simplicity we assume that the mode separation between 
neighboring fibre modes to be $4\pi c/L$, where $c$ denotes the
speed of light. This means that the number of 
fibre modes which couple significantly to the cavity mode
is of the order of $N=\kappa_f L/4\pi c$. We can estimate the 
coupling of the cavity mode to an individual fibre mode by multiplying
$g_f$ by a factor which reflects the discretization of the 
spectrum. From the commutation relations of the field operators 
we conclude that this factor fulfills the following relation:
\[
	f\approx\sqrt{\kappa_f/N}.
\]
Thus the coupling of the cavity mode to an individual fibre mode
is approximately:
\[
\nu = \alpha/\sqrt{L} \approx\sqrt{8\kappa_f\pi c/L}.
\]

As a concrete numerical example let us assume that the decay 
rate of the fibre into the cavity is $\kappa/2\pi=0.5{\rm GHz}$. Thus
we find $\alpha/2\pi\approx0.8{\rm GHz}\sqrt{m}$. Moreover, suppose 
the unwanted cavity loss rate is $\kappa/2\pi=100{\rm MHz}$. In units of
$\alpha$ the cavity loss rate and the mode detuning are 
$\kappa \approx 0.13{\rm m}^{-1/2} \alpha$ and $\Delta_k\approx(k/L)0.77
{\rm m}^{-1/2}\alpha$. We will find later that reliable transfer
can be achieved in principle
for a transfer time 
of the order of $t\approx300{\rm m}^{1/2}\alpha^{-1}\approx 60
{\rm ns}$.
\begin{figure}
\center{\psfig{file=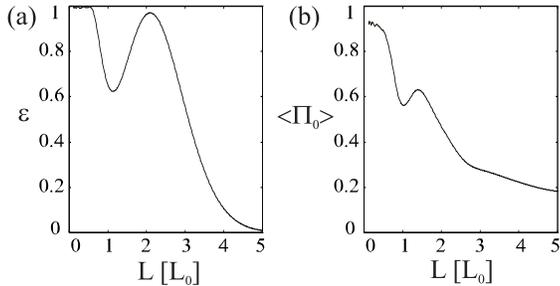,width=2.9in} }
\caption{ (a) Population in the quantum state 
$|a_1,0\rangle |a_0,0\rangle|vac\rangle$$\equiv $$|1\rangle|0\rangle$
after the transfer (initial state 
$|a_0,0\rangle|a_1,0\rangle|vac\rangle$
$\equiv$$|0\rangle|1\rangle$) against the
length of the fibre in units of $L_0$. The total time of the 
transfer was chosen as 
$T$$=$$300$
$\alpha$
$L_0^{1/2}$. 
The pulse shape 
of the two lasers was assumed to be Gaussian: $\Delta_g s_{01}^{A,B}(t)$
$=$
$c{\rm exp}$
$\left(-(t-t_{A,B})^2 /w^2\right)$, with $c=2\alpha L_0^{-1/2}$,
$t_B-t_A=0.2T$ and $w=0.05T$. The scaled cavity loss rate and the separation
between neighbor fibre modes
were chosen as $\kappa=0.1\alpha L_0^{-1/2}$ and $\Delta=0.1
\alpha L_0^{-1/2}$, respectively. 
(b) The time average of the expectation value of the photon number in
the resonant, dark fibre mode divided by the time--average of the
expectation value of the total photon number in the fibre. The parameters
are the same as in (a).
}
\end{figure}
In Fig.~3a the performance of the scheme is studied as a function
of the length of the fibre. 
A good measure is the population $\varepsilon$
in the quantum state $|a_1,0\rangle|a_0,0\rangle|vac\rangle\equiv
|1\rangle|0\rangle$ after the transfer for an inital state 
$|a_0,0\rangle|a_1,0\rangle|vac\rangle\equiv|0\rangle|1\rangle$.
As pointed out earlier the state 
$|a_1,0\rangle|a_1,0\rangle|vac\rangle\equiv|1\rangle|1\rangle$
is decoupled from the laser interaction and thus remains 
unchanged. The length $L$ of the fibre is plotted in units
of $L_0$. For small values of $L$ the transfer is almost perfect.
As the length of the fibre is increased the population $\varepsilon$
decreases (after going through a local maximum) and thus 
the performance of the scheme deteriorates. This behaviour 
is due to the fact that the distance between 
neighbor modes decreases with increasing fibre length. Therefore,
more and more ``grey'' states (i.e. states which are not 
perfectly dark) in the neighborhood of the resonant 
dark state are involved in the transfer. Note that this asymptotic
behaviour is required for causality reasons. If 
all the modes besides the central mode are neglected a transfer between 
the nodes could take place in constant time regardless of the 
length of the fibre. 

Finally, in Fig.~3b I show to what extent the dark state 
Eq.~(\ref{darkstate}) is populated during the transfer. We 
plot $\langle\Pi_0\rangle$, the average population in the dark state 
divided by the average total 
photon number within the fibre. 
Surprisingly, rather good transfer
can be found even in the case where $\langle\Pi_0\rangle$
is not close to one. This suggests that a significant portion of the 
population passes through non--dark states in the neighborhood of 
the dark state and very good transfer can take place for parameters 
where the single mode approximation of the fibre is not at all valid.

In summary, I have proposed a novel scheme to transfer quantum 
states between distant nodes of a quantum network which is 
robust against important sources of decoherence. The scheme
could be used to enable networking of 
several ion trap quantum computers. In addition, error correction
methods could be applied to further increase the stability of
the scheme \cite{vanEnk,Science}. 

I wish to acknowledge fruitful discussions with Artur Ekert,
the members of the quantum information group at Oxford
University, and Jonathan Roberts. 
The author holds
an Erwin--Schr{\"o}dinger scholarship
granted by the Austrian Science Fund.


\vspace{-0.5cm} 

\begin{references}
\bibitem{crypto1}
Zbinden {\it et al.}, Electronics Lett. {\bf 33}, 586 (1997);
C. Marand and P. D. Townsend, Opt. Lett. {\bf 20}, 1695 (1995);
R. J. Hughes {\it et al.}, Contemp. Phys. {\bf 36}, 149 (1995).

\bibitem{crypto2}
S. Wiesner, SIGACT News {\bf 15}, 78 (1983); 
C. H. Bennet and G. Brassard, {\it Proceedings of the IEEE 
Intl. Conf. on Computers,
Systems and Signal Processing}, Bangalore (New York, IEEE, 1984).

\bibitem{qtele}
C. H. Bennet {\it et al.}, Phys. Rev. Lett {\bf 70}, 1895 (1993).

\bibitem{qcomp}
A. Ekert, {\it Proc. 14th International 
    Conference on Atomic Physics ICAP}, ed.
    Smith S., Wieman C., Wineland D., 450, American Institute of Physics, 
    New York (1995); C. Bennett, Phys. Today {\bf 48}(10), 24 (1995);
D. P. DiVincenzo, Science {\bf 270}, 255 (1995); S. Lloyd, Sci. Am. {\bf 273}(4), 44 (1995).

\bibitem{ciraczoller}
J. I. Cirac and  P. Zoller, Phys. Rev. Lett. {\bf 74}, 4091 (1995).

\bibitem{nistboulder}
C. Monroe {\it et al.},
        Phys. Rev. Lett. {\bf 75}, 4714 (1995).

\bibitem{symmetricwavepackets}
J. I. Cirac, P. Zoller, H. J. Kimble, and H. Mabuchi, 
  Phys. Rev. Lett. {\bf 78}, 3221 (1997).

\bibitem{stochschroedinger}
P. Zoller and C. W. Gardiner, in {\it Proceedings
        of the Les Houches Lecture Session LXII: Quantum Fluctuations},
        ed. E. Giacobino and S. Renaud, North Holland, Amsterdam (1996).

\bibitem{quantumnoise}
C. W. Gardiner, {\it Quantum Noise}
        (Springer--Verlag, Berlin 1991).

\bibitem{Ur-AP}
J. Oreg, F. T. Hioe, and J. H. Eberly, Phys. Rev. A {\bf 29}, 690 (1984).

\bibitem{BergmannAP}
B. W. Shore {\it et al.}, Phys. Rev. A {\bf 44}, 7442 (1991).

\bibitem{MarteAP}
P. Marte, P. Zoller, and J. L. Hall, Phys. Rev. A {\bf 44}, R4118 (1991);
L. S. Goldner {\it et al.}, Phys. Rev. Lett. {\bf 72}, 997 (1994);
J. Lawall and M. Prentiss, Phys. Rev. Lett. {\bf 72}, 993 (1994).

\bibitem{ScottAP}
A. S. Parkins {\it et al.}, Phys. Rev. Lett. {\bf 71}, 3095 (1993).

\bibitem{TomAP}
T. Pellizzari, S. A. Gardiner, J. I. Cirac, and P. Zoller,
                Phys. Rev. Lett. {\bf 75}, 3788 (1995).

\bibitem{vanEnk}
S. J. van Enk, J. I. Cirac, and P. Zoller, 
Phys. Rev. Lett. {\bf 78}, 4293 (1997).

\bibitem{Science}
J. I. Cirac, T. Pellizzari and P. Zoller, Science {\bf 273}, 5279
        (1996).


\end{references}
\end{document}